\documentclass[a4paper,10pt]{article}
\usepackage{a4}
\usepackage[dvips]{graphicx}
\usepackage{pdfpages}
\usepackage{epsfig,amsmath,amssymb,verbatim,mathrsfs,array,layout,textcomp,amssymb,latexsym}

\newenvironment{Eqnarray}{\arraycolsep 0.14em\begin{eqnarray}}{\end{eqnarray}}
\def\beqa{\begin{Eqnarray}}
\def\eeqa{\end{Eqnarray}}
\def\beq{\begin{Eqnarray}}
\def\eeq{\end{Eqnarray}}
\newcommand{\no}{\nonumber}

\def\lsim{\mathrel{\rlap{\lower4pt\hbox{\hskip1pt$\sim$}}
     \raise1pt\hbox{$<$}}}         
\def\gsim{\mathrel{\rlap{\lower4pt\hbox{\hskip1pt$\sim$}}
     \raise1pt\hbox{$>$}}}         
\begin{document}
\begin{titlepage}

\vskip1.5cm
\begin{center}
{\Large \bf Model building for flavor changing Higgs couplings}
\end{center}
\vskip0.2cm

\begin{center}
Avital Dery$^1$, Aielet Efrati$^1$, Yosef Nir$^1$, Yotam Soreq$^1$, and Vasja Susi\v{c}$^2$  \\
\end{center}
\vskip 8pt

\begin{center}
{\it $^1${}Department of Particle Physics and Astrophysics,\\
Weizmann Institute of Science, Rehovot 7610001, Israel}\\
{\it $^2${}J. Stefan Institute, 1000 Ljubljana, Slovenia}\\
\vspace*{0.2cm}
{\tt  $^1${}avital.dery,aielet.efrati,yosef.nir,yotam.soreq@weizmann.ac.il} \vspace*{0.2cm}
{\tt $^2${}vasja.susic@ijs.si}
\end{center}

\vskip 8pt

\begin{abstract}
If $t\to hq$ ($q=c,u$) or $h\to\tau\ell$ ($\ell=\mu,e$) decays are observed, it will be a clear signal of new physics. We investigate whether natural and viable flavor models can saturate the present direct upper bounds without violating the indirect constraints from low energy loop processes. We carry out our analysis in two theoretical frameworks: minimal flavor violation~(MFV) and Froggatt-Nielsen symmetry~(FN). The simplest models in either framework predict flavor changing couplings that are too small to be directly observed. Yet, in the MFV framework, it is possible to have  lepton flavor changing Higgs couplings close to the bound if spurions related to heavy singlet neutrinos play a role. In the FN framework, it is possible to have large flavor changing couplings in both the up and the charged lepton sectors if supersymmetry plays a role.
\end{abstract}

\end{titlepage}

\title{\textbf{}}
%


\section{Introduction}
\label{sec:int}
If flavor changing Higgs Yukawa interactions are discovered, it will exclude the Standard Model~(SM) as well as a large class of extensions of its scalar sector. The size and the structure of the flavor changing couplings will probe the electroweak breaking sector, and provide hints about the flavor physics that is responsible for the smallness and hierarchy in the Yukawa parameters.

Experimentally, there are two promising ways to search for flavor changing Higgs couplings. First, one could search for $t\to hq$ decays ($q=c,u$). The present bound on the $htq$ Yukawa couplings reads~\cite{CMS:2014qxa,Aad:2014dya}
\beq\label{eq:htqexp}
\sqrt{|Y_{tq}|^2+|Y_{qt}|^2}\leq0.14.
\eeq
Second, one could search for $h\to\tau^\pm\ell^\mp$ decays ($\ell=\mu,e$)~\cite{Davidson:2012ds,Bressler:2014jta}. The present bound on the $h\tau\mu$ Yukawa coupling reads~\cite{CMS:mutau}
\beq\label{eq:htlexp}
\sqrt{|Y_{\tau\mu}|^2+|Y_{\mu\tau}|^2}\leq0.0036.
\eeq
A direct search for the $h\to\tau e$ decay is also possible~\cite{Bressler:2014jta}, but at present there is only an indirect bound. Assuming that Higgs-mediated loop contributions to $\tau\to\ell\gamma$ decays do not suffer from cancelations against other new physics contributions, the bound on the $h\tau e$ Yukawa couplings reads $\sqrt{|Y_{\tau e}|^2+|Y_{e\tau}|^2}\lsim0.02$~\cite{Harnik:2012pb}.

Experimenters should pursue these searches regardless of whether concrete, well-motivated theoretical models exist which allow for large flavor violation in the Yukawa interactions. Yet, it would be encouraging to know that there exist viable and natural models that allow the bounds in Eqs.~(\ref{eq:htqexp}) and~(\ref{eq:htlexp}) to be saturated. It is the purpose of this work to construct such models, and understand the difficulties involved.

To obtain sizable flavor violation in the light Higgs couplings, we need to extend the SM, either by adding nonrenormalizable terms, or by adding new degrees of freedom, or by both. Whatever framework we consider, we will make the following two assumptions:
\begin{itemize}
\item The scale of the relevant new physics is not much higher than the TeV scale. Otherwise, the flavor violating effects will become unobservably small.
\item The flavor structure of the new physics is not anarchical. Otherwise, TeV scale new physics should have already been manifest in flavor changing neutral current (FCNC) processes.
\end{itemize}
In this work, we examine two such frameworks of flavor physics: minimal flavor violation~(MFV)~\cite{D'Ambrosio:2002ex}, and the Froggatt-Nielsen~(FN) mechanism~\cite{Froggatt:1978nt}.

While the bounds of Eq.~(\ref{eq:htqexp}) are similar for $q=u$ or $c$, and the bound of Eq.~(\ref{eq:htlexp}) for $\ell=\mu$ is stronger than the indirect bound for $\ell=e$, the theoretical frameworks predict larger couplings between the third and second generation fermions than between the third and first. Therefore, we focus on models that saturate the bounds on $Y_{tc}$ and on $Y_{\mu\tau}$.

Before we enter a detailed discussion, we mention several related works. As concerns the $htq$ couplings, Refs.~\cite{Craig:2012vj,Chen:2013qta,Atwood:2013ica,Greljo:2014dka,Wu:2014dba} discuss direct bounds, and Ref.~\cite{Gorbahn:2014sha} discusses indirect ones. Ref.~\cite{Kagan:2014ila} suggests ways to search for flavor changing couplings to light quarks. As concerns the $h\tau\ell$ couplings, Ref.~\cite{Celis:2013xja} suggests ways to use $\tau$ decays to constrain the lepton flavor violating Higgs couplings. For recent  discussions of flavor changing Higgs decays in various other frameworks of new physics, see Refs.~\cite{Bhattacharyya:2010hp,Bhattacharyya:2012ze,Arhrib:2012mg,Arana-Catania:2013xma,Arroyo:2013kaa,Falkowski:2013jya,Arganda:2014dta,Campos:2014zaa}.

\section{Minimal Flavor Violation}
In this section, we study the structure of the Yukawa coupling matrices of the light Higgs boson $h$ under the assumption that it obeys the principle of minimal flavor violation. We identify the flavor suppression of the various flavor changing couplings. Note that a specific framework may introduce additional suppression factors, such as suppression by a high scale of new physics $\Lambda$ (typically, a $v^2/\Lambda^2$ factor) or a loop factor. We ignore such possible additional suppression, bearing in mind that two Higgs doublet models provide an example where the only suppression comes from flavor parameters~\cite{Dery:2013aba}.

We use the notation $y_f$ to denote the SM value of the Yukawa coupling, $y_f=\sqrt2 m_f/v$, where $v\simeq246$ GeV.

\subsection{The up sector}
As concerns the up sector, the largest off-diagonal term is $Y_{ct}$. The largest diagonal term is $Y_{tt}$. The ratio between the two is given by
\beq
\frac{Y_{ct}}{Y_{tt}}=\frac{C^u y_b^2\ V_{cb}V_{tb}^*}{A^u+B^u y_t^2 +C^u y_b^2|V_{tb}|^2}\lsim V_{cb},
\eeq
where $A^f$, $B^f$, $C^f$ are unknown dimensionless coefficients. Note that $C^u$ can be large, up to $C^u y_b^2\sim1$. (One can think of $C^u$ as the MFV analog of the $\tan^2\beta$ factor in models of natural flavor conservation (NFC) of Type II.) The other off-diagonal terms are considerably smaller:
\beqa\label{eq:uctmfv}
\frac{Y_{tc}}{Y_{ct}}&\sim&\frac{V_{tb}V_{cb}^*}{V_{cb}V_{tb}^*}\frac{m_c}{m_t},\no\\
\frac{Y_{ut}}{Y_{ct}}&\sim&\frac{V_{ub}}{V_{cb}},\ \ \
\frac{Y_{tu}}{Y_{ct}}\sim\frac{V_{tb}V_{ub}^*}{V_{cb}V_{tb}^*}\frac{m_u}{m_t},\no\\
\frac{Y_{uc}}{Y_{ct}}&\sim&\frac{V_{ub}V_{cb}^*}{V_{cb}V_{tb}^*}\frac{m_c}{m_t},\ \ \
\frac{Y_{cu}}{Y_{ct}}\sim\frac{V_{ub}^*V_{cb}}{V_{cb}V_{tb}^*}\frac{m_u}{m_t}.
\eeq
Since MFV relates all the off-diagonal couplings to each other, it is important to check that none of the bounds from FCNC would prevent $|Y_{ct}|$ from saturating the direct bound of Eq. (\ref{eq:htqexp}). Indeed, the strongest of the FCNC bounds comes from $D^0-\overline{D}{}^0$ mixing \cite{Blankenburg:2012ex,Harnik:2012pb},
\beq\label{eq:upperyuc}
\sqrt{|Y_{cu}|^2+|Y_{uc}|^2}\leq7\times10^{-5},
\eeq
which is weaker than $0.14(m_c/m_t)|V_{ub}/V_{tb}|$ [see Eqs.~(\ref{eq:htqexp}) and~(\ref{eq:uctmfv})].

We conclude that the only non-negligible coupling is $Y_{ct}$. It is expected to be within the range of $y_b^2|V_{cb}|\lsim |Y_{ct}|\lsim|V_{cb}|$. It can give, at best, a $t\to hc$ decay rate that is an order of magnitude below the present bound. (We note that in the general MFV (GMFV) framework \cite{Kagan:2009bn}, there could be interesting effects on the diagonal coupling $Y_{cc}$~\cite{Delaunay:2013pja}.)

\subsection{The charged lepton sector}
As concerns the charged lepton sector, the implementation of the MFV principle is less straightforward~\cite{Cirigliano:2005ck,Cirigliano:2006su,Davidson:2006bd}. If the neutrinos are Dirac particles, with a Yukawa matrix $Y^\nu$ and eigenvalues $(y_1,y_2,y_3)\propto(m_1,m_2,m_3)$, then
\beqa\label{eq:yijmfv}
\frac{Y_{\mu\tau}}{Y_{\tau\tau}}&=&\frac{C^\nu y_3^2\ U_{\mu3}U_{\tau3}^*}{A^\nu+B^\nu y_\tau^2 +C^\nu y_3^2|U_{\tau3}|^2}\lsim U_{\mu3}/U_{\tau3},\no\\
\frac{Y_{\tau\mu}}{Y_{\mu\tau}}&=&\frac{U_{\tau3}U_{\mu3}^*}{U_{\mu3}U_{\tau3}^*}\frac{m_\mu}{m_\tau},\no\\
\frac{Y_{e\tau}}{Y_{\mu\tau}}&=&\frac{U_{e3}}{U_{\mu3}},\ \ \
\frac{Y_{\tau e}}{Y_{\tau\mu}}=\frac{U_{e3}^*}{U_{\mu3}^*}\frac{m_e}{m_\mu},\no\\
\frac{Y_{e\mu}}{Y_{\mu\tau}}&=&\frac{U_{e3}U_{\mu3}^*}{U_{\mu3}U_{\tau3}^*}\frac{m_\mu}{m_\tau},\ \ \
\frac{Y_{\mu e}}{Y_{\tau\mu}}=\frac{U_{\mu 3}U_{e3}^*}{U_{\tau 3}U_{\mu3}^*}\frac{m_e}{m_\mu}.
\eeqa
Thus, unless the neutrino sector couples to a Higgs doublet with VEV of order $m_3$, the off-diagonal couplings in the charged lepton sector are tiny. In any case, the upper bound from BR$(\mu\to e\gamma)$,
\beq\label{eq:upperyemu}
\sqrt{|Y_{e\mu}|^2+|Y_{\mu e}|^2}\leq1.2\times10^{-6},
\eeq
implies an upper bound on $Y_{\mu\tau}$,
\beq\label{eq:mfvupperymutau}
|Y_{\mu\tau}|\lsim10^{-4},
\eeq
which is two orders of magnitude below the upper bound~(\ref{eq:htlexp}) (as can be seen from Eq.~(\ref{eq:yijmfv}), $Y_{\tau\mu}$ is even smaller by a factor of order $m_\mu/m_\tau$), so that
BR$(h\to\mu\tau)\lsim10^{-4}$BR$(h\to\tau\tau)$. Thus, for lepton MFV with Dirac neutrinos, the rate for $h\to\mu\tau$ is unobservably small.

A more plausible minimal lepton flavor violation (MLFV) scenario is one where the neutrinos are Majorana particles, with their masses generated by the seesaw mechanism. If the seesaw scale is above the scale of flavor dynamics, then the lepton flavor symmetry is $SU(3)_L\times SU(3)_E$, and it is broken by the charged lepton Yukawa matrix $Y^e(\bar3,3)$  only. In this case, there are no lepton flavor changing couplings. If the seesaw scale is below the scale of flavor dynamics, then the lepton flavor symmetry is $SU(3)_L\times SU(3)_E\times SU(3)_N$, and it is broken by three spurions: $\hat Y^e(\bar3,3,1)$, $\hat Y^\nu(\bar3,1,3)$ and $\hat Z^N(1,1,6)$.

In the charged lepton and heavy neutrino mass basis, we can choose our spurions as follows:
\beqa
\hat Y^e&=&Y^e_M\equiv{\rm diag}(y_e,y_\mu,y_\tau)=(\sqrt2/v){\rm diag}(m_e,m_\mu,m_\tau),\no\\
\hat Z^N&=&Z^N_M\equiv{\rm diag}(M_1,M_2,M_3)/m_N,
\eeq
where $M_i$ are the heavy neutrino masses and $m_N$ is their mass scale.
In this basis, the Yukawa matrix $\hat Y^\nu$ assumes the form~\cite{Casas:2001sr}
\beq
\hat Y^\nu=i\frac{\sqrt{m_N}}{v}\sqrt{Z^N_M}R\sqrt{m_\nu}U^\dagger,
\eeq
where $m_\nu={\rm diag}(m_1,m_2,m_3)$, $U$ is the leptonic mixing matrix, and $R$ is a general orthogonal matrix.

The leading contribution to flavor changing Higgs couplings is of the form
\beq\label{eq:yeij}
\frac{Y^e_{ij}}{y_j}\propto (\hat Y^{\nu\dagger}\hat Y^\nu)_{ij}=
\frac{m_N}{v^2}(U\sqrt{m_\nu}R^\dagger Z^N_M R\sqrt{m_\nu}U^\dagger)_{ij}.
\eeq
Note that, unlike the quark sector, there are here several unknowns. Concretely, $R$ and $Z^N_M$ are completely unknown, while there is only partial information on $m_\nu$.

\subsubsection{Degenerate heavy neutrinos}
To start with a simple and predictive model, we make the following simplifying assumptions:
\begin{enumerate}
\item The heavy neutrinos are degenerate: $Z^N_M\propto{\bf 1}$;
\item The matrix $R$ is real.
\end{enumerate}
Then, Eq.~(\ref{eq:yeij}) simplifies into
\beq\label{eq:yeija}
\frac{Y^e_{ij}}{y_j}\propto\frac{m_N}{v^2}(U m_\nu U^\dagger)_{ij}.
\eeq
We can now consider the three classes of light neutrino spectrum:

(i) Degeneracy, $m_1\simeq m_2\simeq m_3$: In the degenerate limit, the unitarity of $U$ gives
\beq
Y^e_{ij}=0.
\eeq

(ii) Inverted hierarchy, $m_3\ll m_1\simeq m_2$:
Taking the limit of $m_3=0$ and defining, for convenience, $m_N=v^2/m_2$, we obtain
\beq
\frac{Y^e_{ij}}{y_j}\propto U_{i3}U_{j3}^*.
\eeq
In particular, we have
\beq\label{eq:yeijih}
\left|\frac{Y^e_{e\mu}}{Y^e_{\mu\tau}}\right|=\left|\frac{U_{e3}}{U_{\tau3}}\right|\frac{m_\mu}{m_\tau}\sim10^{-2}.
\eeq

(iii) Normal hierarchy, $m_1\ll m_2\ll m_3$:
Taking the limit of $m_1=0$ and defining, for convenience, $m_N=v^2/m_3$, we obtain
\beq
\frac{Y^e_{ij}}{y_j}\propto (m_2/m_3)U_{i2}U_{j2}^*+U_{i3}U_{j3}^*.
\eeq
In particular, we have
\beq\label{eq:yeijnh}
\left|\frac{Y^e_{e\mu}}{Y^e_{\mu\tau}}\right|=
\left|\frac{(m_2/m_3)U_{e2}U_{\mu2}^*+U_{e3}U_{\mu3}^*}{(m_2/m_3)U_{\mu2}U_{\tau2}^*+U_{\mu3}U_{\tau3}^*}\right|
\frac{m_\mu}{m_\tau}\sim10^{-2}.
\eeq

In both cases of inverted hierarchy, Eq.~(\ref{eq:yeijih}), and normal hierarchy, Eq.~(\ref{eq:yeijnh}), the upper bound of Eq.~(\ref{eq:upperyemu}) implies an upper bound on $Y_{\mu\tau}$ similar to Eq.~(\ref{eq:mfvupperymutau}). The conclusion is that an observably large value of $Y_{\mu\tau}$ in MLFV models requires that $Z^N_M$ and $R$ play a non-trivial role in the structure of $Y^\nu$, such that the ratio $|Y_{e\mu}/Y_{\mu\tau}|$ is much smaller than the values of order $10^{-2}$ obtained in the simplest models.

\subsubsection{Hierarchical heavy neutrinos}
Our starting point is, again, Eq.~(\ref{eq:yeij}). We relax the assumption of degenerate heavy neutrinos, but for the sake of simplicity keep $R$ real. For off-diagonal couplings, we can write
\beq
Y_{ij}/y_j\propto A_{kl}U_{ik}U_{jl}^*\ {\rm where}\ A\equiv\sqrt{m_\nu}R^T Z_M^N R\sqrt{m_\nu} \, .
\eeq
In particular,
\beqa
Y_{e\mu}&\propto&U_{ei}A_{ij}U_{\mu j}^*=\sum_i A_{ii}U_{ei}U_{\mu i}^*
+\sum_{i<j}A_{ij}(U_{ei}U_{\mu j}^*+U_{ej}U_{\mu i}^*),\no\\
Y_{\mu\tau}&\propto&U_{\mu i}A_{ij}U_{\tau j}^*=\sum_i A_{ii}U_{\mu i}U_{\tau i}^*
+\sum_{i<j}A_{ij}(U_{\mu i}U_{\tau j}^*+U_{\mu j}U_{\tau i}^*).
\eeqa
Since we aim to construct a model where the hierarchy between $Y_{e\mu}$ and $Y_{\mu\tau}$ is much stronger than in the simple models of the previous subsection, we try to obtain $Y_{e\mu}=0$, while $Y_{\mu\tau}\neq0$. The first sum on the right hand side vanishes only when the three $A_{ii}$ are equal. This ansatz leads, however, to the vanishing of the first sum in the expression for $Y_{\mu\tau}$ as well. As concerns the second sum, requiring that it vanishes for $Y_{e\mu}$ is equivalent to solving a complex equation with two real unknowns:
\beq
(U_{e1}U_{\mu2}^*+U_{e2}U_{\mu1}^*)+\frac{A_{23}}{A_{12}}(U_{e2}U_{\mu3}^*+U_{e3}U_{\mu2}^*)
+\frac{A_{13}}{A_{12}}(U_{e1}U_{\mu3}^*+U_{e3}U_{\mu1}^*)=0.
\eeq
This equation has a single solution for the two independent unknowns, $A_{23}/A_{12}$ and $A_{13}/A_{12}$. Choosing these values, the second sum in the expression for $Y_{e\mu}$ vanishes, but the corresponding sum in $Y_{\mu\tau}$ does not. In particular, we can choose the entries of $R$ and $M_N$ in such a way that $Y_{e\mu}=0$, while $Y_{\mu\tau}\sim0.02$.

\subsection{Summary}
Our conclusions concerning the $htq$ and $h\tau\ell$ couplings in the MFV framework are the following:
\begin{enumerate}
\item The $Y_{ct}$ coupling is small, with flavor suppression in the range $(y_b^2|V_{cb}|,|V_{cb}|)$. The $Y_{tc},Y_{ut}$ and $Y_{tu}$ couplings are even smaller. Thus, BR$(h\to tq)$ is at least one order of magnitude below present bounds.
\item In the simplest models of MLFV, with the charged lepton Yukawa matrix as a single spurion, the $Y_{\tau\ell}$ and $Y_{\ell\tau}$ couplings vanish.
\item If the unknown dimensionless coefficients in the MLFV expansion do not fulfill special relations, and if the Higgs-mediated contributions to $\mu\to e\gamma$ do not cancel against other new physics contributions, then the upper bounds on $|Y_{e\mu}|$ and $|Y_{\mu e}|$ imply that BR$(h\to\tau\ell)$ is at least four orders of magnitude below BR$(h\to\tau\tau)$.
\item For specific relations between the dimensionless coefficients, it is possible that $Y_{e\mu}$ is accidentally suppressed while $Y_{\mu\tau}$ saturates the upper bound from $\tau\to\mu\gamma$. In this case, BR$(h\to\tau\ell)$ can be comparable to BR$(h\to\tau\tau)$.
\end{enumerate}

\section{Froggatt-Nielsen symmetry}
The Froggatt-Nielsen mechanism~\cite{Froggatt:1978nt} provides a simple explanation for the smallness and hierarchy in the flavor parameters. Selection rules that follow from an approximate Abelian symmetry imply that the various Yukawa couplings are suppressed by different powers of the small symmetry breaking parameter.

When we need to be specific, we employ as the FN symmetry a $[U(1)]^n$ symmetry broken by spurions $\epsilon_i$ of charge $-1$ under the $U(1)_i$ and $0$ under all $U(1)_{j\neq i}$ in the FN symmetry, and numerical value similar to the Cabibbo angle, $\epsilon_i\simeq\lambda=0.2$. Our models should give the estimated parametric suppression of the measured mass and mixing parameters, which we present in Table~\ref{tab:massmix}, as well as satisfy the bounds on the various flavor changing Higgs couplings, which we present in Table~\ref{tab:fcbounds}.

In this section, the ``$\sim$" sign stands for ``has the same parametric suppression as". When it appears for a matrix, it means that each entry in the matrix has an independent and unknown order one coefficient which we do not write explicitly.

\begin{table}[t]
\caption{Our estimates of the parametric suppression of the SM flavor parameters in terms of $\lambda=0.2$. Here, $y_f=\sqrt2 m_f/v$.}
\label{tab:massmix}
\begin{center}
\begin{tabular}{cc} \hline\hline
\rule{0pt}{1.2em}%
parameters & suppression  \\[2pt] \hline\hline
\rule{0pt}{1.2em}%
$y_u,y_c,y_t$ & $\lambda^7,\lambda^3,1$  \\
$y_d,y_s,y_b$ & $\lambda^6,\lambda^4,\lambda^2$  \\
$y_e,y_\mu,y_\tau$ & $\lambda^8,\lambda^5,\lambda^3$  \\
$|V_{us}|,|V_{cb}|,|V_{ub}|$ & $\lambda,\lambda^2,\lambda^3$ \\
$|U_{e2}|,|U_{\mu3}|,|U_{e3}|$ & $1,1,1$ \\
\hline\hline
\end{tabular}
\end{center}
\end{table}

\begin{table}[t]
\caption{Upper bounds on flavor changing Higgs couplings in terms of $\lambda=0.2$.}
\label{tab:fcbounds}
\begin{center}
\begin{tabular}{cc} \hline\hline
\rule{0pt}{1.2em}%
coupling & upper bound  \\[2pt] \hline\hline
\rule{0pt}{1.2em}%
$Y_{uc},Y_{ct},Y_{ut}$ & $\lambda^6,\lambda,\lambda$  \\
$Y_{ds},Y_{sb},Y_{db}$ & $\lambda^8,\lambda^5,\lambda^4$  \\
$Y_{e\mu},Y_{\mu\tau},Y_{e\tau}$ & $\lambda^8,\lambda^3,\lambda^3$  \\
\hline\hline
\end{tabular}
\end{center}
\end{table}

\subsection{The SM with nonrenormalizable terms}
 As concerns the Yukawa couplings of the Higgs, the simplest extension of the SM where we can study modifications that are subject to the FN selection rules is the addition of nonrenormalizable terms. Concretely, in addition to the SM Yukawa terms,
\beq
{\cal L}_Y=-\lambda_{ij}^uQ_i\bar U_j\phi-\lambda_{ij}^dQ_i\bar D_j\phi^\dagger
-\lambda_{ij}^eL_i\bar E_j\phi^\dagger+{\rm h.c.},
\eeq
we consider the dimension-six terms
\beq
{\cal L}_Y^{d=6}=-\frac{\lambda_{ij}^{\prime u}}{\Lambda^2}Q_i\bar U_j\phi(\phi^\dagger\phi)
-\frac{\lambda_{ij}^{\prime d}}{\Lambda^2}Q_i\bar D_j\phi^\dagger(\phi^\dagger\phi)
-\frac{\lambda_{ij}^{\prime e}}{\Lambda^2}L_i\bar E_j\phi^\dagger(\phi^\dagger\phi)
+{\rm h.c.}.
\eeq
Defining the unitary matrices $V^f_L$ and $V^f_R$ via
\beq
\sqrt2 m^f=V^f_L\left(\lambda^f+\frac{v^2}{2\Lambda^2}\lambda^{\prime f}\right)V_R^{f\dagger}v,
\eeq
where $m^f$ is the diagonal mass matrix,  and defining $\hat\lambda^f$ via
\beq
\hat\lambda^f=V_L^f\lambda^{\prime f}V_R^{f\dagger},
\eeq
we obtain for the Yukawa matrix in the fermion mass basis
\beq
Y^f_{ij}=\frac{\sqrt2 m^f_i}{v}\delta_{ij}+\frac{v^2}{\Lambda^2}\hat\lambda^f_{ij}.
\eeq

Imposing the FN mechanism on the $\lambda$ and $\lambda^\prime$ matrices, we obtain the following form for the $\hat\lambda$ matrices~\cite{Dery:2013rta}:
\beqa
\hat\lambda^u&\sim&\begin{pmatrix}
y_u & |V_{us}|y_c & |V_{ub}|y_t \\ y_u/|V_{us}| & y_c & |V_{cb}|y_t \\
y_u/|V_{ub}| & y_c/|V_{cb}| & y_t \end{pmatrix},\no\\
\hat\lambda^d&\sim&\begin{pmatrix}
y_d & |V_{us}|y_s & |V_{ub}|y_b \\ y_d/|V_{us}| & y_s & |V_{cb}|y_b \\ y_d/|V_{ub}| & y_s/|V_{cb}| & y_b \end{pmatrix},\no\\
\hat\lambda^e&\sim&\begin{pmatrix}
y_e & |U_{e2}|y_\mu & |U_{e3}|y_\tau \\ y_e/|U_{e2}| & y_\mu & |U_{\mu3}|y_\tau \\ y_e/|U_{e3}| & y_\mu/|U_{\mu3}| & y_\tau \end{pmatrix}.
\eeqa

These predictions have to be compared with the phenomenological upper bounds on the various entries in the $Y^f$ matrices, Eq.~(\ref{eq:upperyuc}) for the up sector, Eq.~(\ref{eq:upperyemu}) for the charged lepton sector, and, for the down sector~\cite{Harnik:2012pb,Blankenburg:2012ex},
\beqa\label{eq:upperyds}
\sqrt{|Y_{ds}|^2+|Y_{sd}|^2}&\leq&2\times10^{-5},\no\\
\sqrt{|Y_{db}|^2+|Y_{bd}|^2}&\leq&2\times10^{-4},\no\\
\sqrt{|Y_{sb}|^2+|Y_{bs}|^2}&\leq&1\times10^{-3}.
\eeqa
These bounds require that
\beq
v^2/\Lambda^2\lsim10^{-2},
\eeq
rendering the $Y_{qt}$ and $Y_{\ell\tau}$ couplings unobservably small.

This result cannot be circumvented without supersymmetry. The reason is that the parametric suppression of the $\lambda^f$ matrices is fully dictated by the fermion masses and mixing, and that, since $\phi^\dagger\phi$ does not carry charge, the parametric suppression of $\lambda^{\prime f}$ is the same as that of $\lambda^f$. Thus, to make progress in our model building, we need to incorporate supersymmetry.

\subsection{The MSSM with nonrenormalizable terms}
Working in a supersymmetric framework opens up new possibilities for flavor model building. In particular, the requirement that the superpotential is holomorphic can lead to interesting consequences:
\begin{enumerate}
\item Holomorphicity does not allow for $\phi_q^\dagger \phi_q$ factors in the superpotential. Thus, the relevant higher order terms include a $\phi_u\phi_d$ factor.
\item The fact that $\phi_u\phi_d$ may carry a FN charge implies that the structure of $\lambda^f$ and $\lambda^{\prime f}$ is not necessarily the same.
\item If a term in the superpotential carries charge of the same sign as the relevant spurion, this term vanishes. This situation is known as `holomorphic zero'~\cite{Leurer:1993gy}.
\end{enumerate}

We consider the following terms in the superpotential:
\beqa\label{eq:wmssmnr}
W_Y&=&\lambda_{ij}^u Q_i \bar U_j\phi_u-\lambda_{ij}^d Q_i \bar D_j\phi_d-\lambda_{ij}^e L_i \bar E_j\phi_d\\
&+&\frac{\lambda_{ij}^{\prime u}}{\Lambda^2} Q_i \bar U_j\phi_u(\phi_u\phi_d)
-\frac{\lambda_{ij}^{\prime d}}{\Lambda^2} Q_i \bar D_j\phi_d(\phi_u\phi_d)
-\frac{\lambda_{ij}^{\prime e}}{\Lambda^2} L_i \bar E_j\phi_d(\phi_u\phi_d).\no
\eeqa
The charged fermion mass matrices are given by
\beqa
\sqrt2 m^u&=&V_L^u\left[\lambda^u+\frac{v_uv_d}{2\Lambda^2}\lambda^{\prime u}\right]V_R^{u\dagger} v_u,\no\\
\sqrt2 m^d&=&V_L^d\left[\lambda^d+\frac{v_uv_d}{2\Lambda^2}\lambda^{\prime d}\right]V_R^{d\dagger} v_d,\no\\
\sqrt2 m^e&=&V_L^e\left[\lambda^e+\frac{v_uv_d}{2\Lambda^2}\lambda^{\prime e}\right]V_R^{e\dagger} v_d.
\eeqa
The Yukawa matrices of $\phi_u$ and $\phi_d$ in the fermion mass basis are given by
\beqa
(Y_{\phi_u}^u)_{ij}&=&\frac{m^u_i}{v_u}\delta_{ij}+\frac{v_u v_d}{2\sqrt2\Lambda^2}(\hat\lambda^u)_{ij},\ \ \
(Y_{\phi_d}^u)_{ij}=\frac{v_u^2}{2\sqrt2\Lambda^2}(\hat\lambda^u)_{ij},\no\\
(Y_{\phi_d}^d)_{ij}&=&\frac{m^d_i}{v_d}\delta_{ij}+\frac{v_u v_d}{2\sqrt2\Lambda^2}(\hat\lambda^d)_{ij},\ \ \
(Y_{\phi_u}^d)_{ij}=\frac{v_d^2}{2\sqrt2\Lambda^2}(\hat\lambda^d)_{ij},\no\\
(Y_{\phi_d}^e)_{ij}&=&\frac{m^e_i}{v_d}\delta_{ij}+\frac{v_u v_d}{2\sqrt2\Lambda^2}(\hat\lambda^e)_{ij},\ \ \
(Y_{\phi_u}^e)_{ij}=\frac{v_d^2}{2\sqrt2\Lambda^2}(\hat\lambda^e)_{ij}.
\eeqa
Defining an angle $\beta$ via $\tan\beta \equiv v_u/v_d$, an angle $\alpha$ via
\beq
h=-s_\alpha{\cal R}e(\phi_d^0)+c_\alpha {\cal R}e(\phi_u^0),
\eeq
and
\beq
\kappa\equiv\frac{v^2}{2\sqrt2\Lambda^2}\cos(\alpha+\beta),
\eeqa
we have
\beqa
Y_h^u&=&+(c_\alpha/s_\beta)(m^u/v)+\kappa s_\beta\hat\lambda^u,\no\\
Y_h^d&=&-(s_\alpha/c_\beta)(m^d/v)+\kappa c_\beta\hat\lambda^d,\no\\
Y_h^e&=&-(s_\alpha/c_\beta)(m^e/v)+\kappa c_\beta\hat\lambda^e.
\eeqa
The first terms on the right hand side of the equations for $Y_h^u$, $Y_h^d$ and $Y_h^e$ are the well-known, flavor-diagonal, expressions for type II 2HDM models. The second terms are the new, flavor-changing, contributions due to the nonrenormalizable terms.

We run into two potentially conflicting requirements (from here on we omit the sub-index $h$ for the Yukawa couplings of the light Higgs, and the super-index $u,d,e$ wherever unnecessary):
\begin{enumerate}
\item We aim to have $|Y_{tc}|$ ($|Y_{\mu\tau}|$) large enough to allow for observable $t\to ch$ ($h\to\tau\mu$) decay rate.
\item The flavor changing $|Y^f_{ij}|$ couplings have to be small enough to obey the phenomenological bounds of Table~\ref{tab:fcbounds}.
\end{enumerate}
Let us elaborate on these points, and find the general consequences for model building.

As concerns the up sector, we would like to have $|Y_{tc}|=\kappa s_\beta\hat\lambda^u_{32}$ of order 0.2. Since $\kappa$ by itself induces such, or stronger, suppression, we need to have $\hat\lambda^u_{32}$ unsuppressed. This, in turn, requires that we have $\lambda^{\prime u}_{32}={\cal O}(1)$. Given the bounds on other off-diagonal couplings in $Y^u$, we conclude that all other off-diagonal entries in $\lambda^{\prime u}$ must be holomorphic zeros. Furthermore, taking into account that the rotation from $\lambda^{\prime u}$ to $\hat\lambda^u$ depends on the structure of $\lambda^u$, we learn that $\lambda^u_{13}$, and either $\lambda^u_{23}$ or $\lambda^u_{12}$, must be holomorphic zeros.

As concerns the down sector, the phenomenological constraints on the off-diagonal couplings are very stringent. Since we do not have as one of our goals making any of them large, we employ the simplest (though not the only) strategy for our model building and that is to have $\lambda^{\prime d}=0$, which avoids all relevant constraints.

\subsection{A $U(1)\times U(1)$ Model}
We consider a model with a $U(1)_1\times U(1)_2$ FN symmetry. The symmetry is broken by two spurions:
\beq
\epsilon_1(-1,0)\sim\lambda,\ \ \ \epsilon_2(0,-1)\sim\lambda.
\eeq
The two Higgs doublets are assigned the following FN charges:
\beq
\phi_u(+1,-1),\ \ \ \ \phi_d(+0,-1).
\eeq
The three quark generations are assigned the following FN charges:
\beqa
&&Q_1(+4,-1),\ \ \ Q_2(+3,-1),\ \ \ Q_3(0,0),\no\\
&&\bar D_1(+2,+2),\ \ \ \bar D_2(+1,+2),\ \ \ \bar D_3(+1,+2),\no\\
&&\bar U_1(+2,+2),\ \ \ \bar U_2(-2,+3),\ \ \ \bar U_2(-1,+1).
\eeqa
The three lepton generations are assigned the following FN charges:
\beqa
&&L_1(+2,+1),\ \ \ L_2(-1,+4),\ \ \ L_3(0,+3),\no\\
&&\bar E_1(-2,+8),\ \ \ \bar E_2(+5,-2),\ \ \ \bar E_3(+2,-1).
\eeqa

We get the following parametric suppressions and holomorphic zeros in the $\lambda$ matrices:
\beq
\lambda^u\sim
\begin{pmatrix}
\lambda^{7}&\lambda^{4}&0\\
\lambda^{6}&\lambda^{3}&0\\
\lambda^{4}&0&1\\
\end{pmatrix},\ \ \
\lambda^d\sim
\begin{pmatrix}
\lambda^{6}&\lambda^{5}&\lambda^{5}\\
\lambda^{5}&\lambda^{4}&\lambda^{4}\\
\lambda^{3}&\lambda^{2}&\lambda^{2}\\
\end{pmatrix},\ \ \
\lambda^e\sim
\begin{pmatrix}
\lambda^8&0&0\\
0&\lambda^5&\lambda^3\\
0&\lambda^5&\lambda^3\\
\end{pmatrix}.
\eeq
The matrices $\lambda^u$ and $\lambda^d$ generate the quark mass eigenvalues and the CKM mixing angles with the parametric suppression given in Table~\ref{tab:massmix}. The matrix $\lambda^e$ generates the charged lepton mass eigenvalues of Table~\ref{tab:massmix}. As concerns the lepton mixing angles, one has to take into consideration also the neutrino mass matrix. We do not present here explicitly the resulting neutrino mass matrices, as there are subtleties in the interplay between the FN mechanism and the seesaw mechanism~\cite{Nir:2004my}. The charge assignments of the $L_i$ fields imply, however, that the flavor structure of $\lambda^\nu$, which appears in the dimension five terms $(\lambda^\nu_{ij}/\Lambda_L)L_iL_j\phi_u\phi_u$, with $\Lambda_L$ the seesaw scale, can be anarchical, thus providing the required large leptonic mixing angles.

We get the following parametric suppressions and holomorphic zeros in the $\lambda^\prime$ matrices:
\beq
\lambda'^u\sim \frac{1}{\lambda}
\begin{pmatrix}
0&0&0\\
0&0&0\\
0&\lambda&0\\
\end{pmatrix},\ \ \
\lambda'^d=0,\ \ \
\lambda'^e\sim \frac{1}{\lambda}
\begin{pmatrix}
\lambda^8&0&0\\
0&0&\lambda^3\\
0&0&0\\
\end{pmatrix}.
\eeq

The resulting $\hat\lambda$ matrices have the following form:
\beq\label{eq:hatlambdamodel}
\hat{\lambda}^u\sim \frac{1}{\lambda}
\begin{pmatrix}
\lambda^{15}&\lambda^{12}&\lambda^{13}\\
\lambda^{8}&\lambda^{5}&\lambda^{6}\\
\lambda^{4}&\lambda^{1}&\lambda^{2}\\
\end{pmatrix},\ \ \
\hat{\lambda}^d=0,\ \ \
\hat{\lambda}^e\sim \frac{1}{\lambda}
\begin{pmatrix}
\lambda^8&0&0\\
0&\lambda^5&\lambda^3\\
0&\lambda^5&\lambda^3\\
\end{pmatrix}.
\eeq
Thus, the couplings of interest can saturate the present bounds,
\beqa
|Y_{tc}|&\sim&\kappa s_\beta,\no\\
|Y_{\mu\tau}|&\sim&\kappa c_\beta \lambda^2.
\eeqa

As concerns the other flavor changing couplings, all of them satisfy the bounds quoted in Table~\ref{tab:fcbounds}. As concerns the diagonal couplings, the $\hat\lambda$ terms may lead to a violation of the SM relation between the mass and the Yukawa coupling. Within our specific model, Eq.~(\ref{eq:hatlambdamodel}) implies that there are no such modifications in the down sector and there are small deviations, of ${\cal O}(\lambda^2)$, in the up sector. In the charged lepton sector, however, the deviations are of order one. Our model provides then an explicit example of how measurements of the Higgs decay rates into $\tau^+\tau^-$, $\mu^+\mu^-$ and $\tau^\pm\mu^\mp$ can probe flavor models, as envisioned in Ref.~\cite{Dery:2013rta}.

\subsection{Summary}
Our conclusions concerning the $htq$ and $h\tau\ell$ couplings in the FN framework are the following:
\begin{enumerate}
\item Within the SM with non-renormalizable terms, the FCNC bounds imply that all flavor changing couplings are too small for direct observation.
\item Within the MSSM with non-renormalizable terms, it is possible to construct models such that the $htc$ and $h\tau\mu$ couplings are close to present bounds and all FCNC bounds are satisfied.
\item The models that achieve our goals are not generic. Specifically, the FN charges are very restricted.
\item Our models allow not only flavor changing couplings, but also modifications of flavor diagonal ones.
\end{enumerate}
We further note the following points regarding model building in the framework of the MSSM with non-renormalizable terms and FN selection rules:
\begin{itemize}
\item Models with a single $U(1)$ as the FN symmetry cannot achieve our goals. Yet, it is interesting to note that even a single $U(1)$ makes it possible to forbid all the corrections to the Yukawa terms from non-renormalizable terms.
\item The $U(1)\times U(1)$ model that we presented in this section is not unique. We chose to present it because it involves the smallest FN charges among the models that we found.
\item Models with a $[U(1)]^3$ FN symmetry are much less restrictive than $[U(1)]^2$ models.
\item Models where the combination $(\phi_u\phi_d)$ carries FN charges have implications for the $\mu$ and the $B$ terms. In particular, they provide a possible explanation of why $\mu$ is close to the electroweak scale rather than to the Planck scale~\cite{Nir:1995bu} via the Giudice-Masiero mechanism~\cite{Giudice:1988yz}.
\end{itemize}

\section{The Charged Higgs} \label{sec:Charged}
The MSSM predicts the existence of five physical Higgs scalars, including a charged Higgs, $H^\pm$. Within this model, the tree level couplings of the charged Higgs depend only on the fermions masses, the CKM parameters, and $\tan\beta$. In particular, they are independent of the soft supersymmetry breaking terms. The experimental searches for the supersymmetric charged Higgs are based on these predictions. In the previous section we learned that, within the MSSM as a low energy effective theory, the Yukawa couplings of the light CP-even neutral scalar are modified in various interesting ways. It is then interesting to understand whether similarly significant modifications might occur in this framework for the charged Higgs.

Our starting point is, again, the superpotential of Eq.~(\ref{eq:wmssmnr}). In the quark mass basis, we define the Yukawa matrices for the charged scalars, $h_u^+$ and $h_d^-$, as follows:
\beq
{\cal L}_Y\supset \frac{(Y_{h_u^+})_{ij}}{\sqrt2}\overline{u_{Ri}}d_{Lj}h_u^+
+\frac{(Y_{h_d^-})_{ij}}{\sqrt2}\overline{d_{Ri}}u_{Lj}h_d^-+{\rm h.c.}.
\eeq
We obtain:
\beqa
Y_{h^+_u}&=&V_R^u\left(\lambda^u+\frac{v_uv_d}{2\Lambda^2}\lambda^{\prime u}\right)V_L^{d\dagger}=\frac{\sqrt{2}m_u}{v_u}V\,,\no\\
Y_{h^-_d}&=&V_R^d\left(\lambda^d+\frac{v_uv_d}{2\Lambda^2}\lambda^{\prime d}\right)V_L^{u\dagger}=\frac{\sqrt{2}m_d}{v_d}V^\dagger,
\eeqa
where $V$ is the CKM matrix.
Thus, the presence of the non-renormalizable $\lambda^\prime$ terms does not change the relation between the Yukawa couplings of the charged Higgs and the quark mass and mixing parameters. The source of this difference between $h$ and $H^\pm$ is that for the charged Higgs there is no combinatorial factor that is different between the $\lambda^\prime$ contributions to the mass matrix and to the Yukawa matrix.

For the charged Higgs mass eigenstate, $H^+=c_\beta h_u^+-s_\beta h_d^{-*}$, the tree level decay rate into a quark pair,
\beq
\Gamma(H^+\to u_i\overline{d_j})=\frac{N_c G_F m_{H^+}}{4\sqrt2\pi}|V_{ij}|^2
\left[m_{u_i}^2\cot^2\beta+m_{d_j}^2\tan^2\beta\right],
\eeq
is the same as in the renormalizable MSSM.

\section{Conclusions}
The experimental effort to measure the Yukawa couplings of the newly discovered scalar $h$ focusses on the third generation couplings, $Y_t$, $Y_b$ and $Y_\tau$, where the Standard Model predicts large enough rates to be measured. It is possible, however, that new physics which affects only little these leading couplings will enhance couplings to the lighter generations in a more significant way, and/or generate off-diagonal couplings (and even generate CP violation in these off-diagonal couplings~\cite{Kopp:2014rva}). Experimenters should search for such signals of new physics regardless of theoretical prejudices. Yet, it is interesting to investigate whether viable, natural and well-motivated models of new physics can accommodate observably large effects. The goal of this work is to answer this question.

We focussed on models where the scale of new physics is low enough such that  $|Y_{tq}|={\cal O}(0.1)$ and/or $|Y_{\tau\ell}|={\cal O}(0.003)$ can be generated in principle. We further required that the flavor structure of the new physics is not arbitrary, but rather dictated by a dynamical or approximate symmetry principle. Concretely, we employed two frameworks, which well demonstrate the
range of possibilities within natural flavor models. On one side, minimal flavor violation (MFV) requires that the flavor structure of new physics is essentially the same as that of the Standard Model. On the other side, the Froggatt-Nielsen framework requires only that the new physics and the Standard Model share the same parametric suppressions, following from selection rules, but allows all order one coefficients to be unrelated.

What is common to almost all flavor models is that they relate, in either an exact or an approximate way, all off-diagonal couplings within a given fermion sector. If such a connection exists, then the bounds from $\mu\to e\gamma$ imply that all lepton flavor violating effects are too small for direct observation. We found however ways to avoid this relation in both frameworks. In the leptonic MFV framework, we exploit the unmeasured seesaw parameters to suppress $Y_{e\mu}$ compared to $Y_{\mu\tau}$. In the FN framework, we use holomorphic zeros that can arise in Supersymmetry to achieve this effect.

As concerns the up sector, within MFV, the bounds from $D^0-\overline{D}{}^0$ mixing do not prevent a large $Y_{ct}$ coupling. However, the direct MFV constraint on this coupling implies that $|Y_{ct}|\lsim|V_{cb}|$ and perhaps 2-3 orders of magnitude below this upper bound. In other words, if $t\to hq$ is observed very close to the present experimental upper bound, the MFV framework will be excluded. Within the FN framework, the $Y_{uc}-Y_{tc}$ relation does pose a problem, and we again have to generate holomorphic zeros to undo the relation.

Our conclusion is that the upper bounds on the flavor changing Higgs couplings involving the top quark, and/or the tau lepton can be saturated within viable and natural flavor models. These models are, however, not generic, and the careful selection of models within the generic frameworks has no special motivation. If $t\to hq$ or $h\to\tau\ell$ is observed in experiments, it will challenge present explanations of the flavor puzzles.

\vspace{1cm}
\begin{center}
{\bf Acknowledgements}
\end{center}
YN is the Amos de-Shalit chair of theoretical physics. YN is supported by the Israel Science Foundation, and by the I-CORE program of the Planning and Budgeting Committee and the Israel Science Foundation (grant number 1937/12). VS would like to thank the Weizmann Institute for its hospitality and support during his stay there in Spring 2014, when part of this work was done. VS is supported by the Slovenian Research Agency.


\end{document}